\documentclass[10pt]{article}

\pdfoutput=1
\usepackage[T1]{fontenc}
\usepackage[latin1]{inputenc}
\usepackage{array}
\usepackage{subfigure}
\usepackage{graphicx}
\usepackage{multirow}
\usepackage[hmargin=3cm,vmargin=3.5 cm]{geometry}



\begin{document}
\title{On the statistical significance of temporal firing patterns in multi-neuronal spike trains}

\author{
Casey Diekman\footnote{Dept. Industrial \& Operations Engineering, Center for Computational Medicine \& Biology, University
of Michigan, Ann Arbor, MI 48109, USA. Email: diekman@umich.edu},  
P.S. Sastry\footnote{Dept. Electrical Engineering, Indian Institute of Science,
Bangalore 560012, India. Email: sastry@ee.iisc.ernet.in} and K.P. Unnikrishnan\footnote{General Motors R\&D Center, Warren, MI 48090, USA. Email: k.unnikrishnan@gm.com }
 }

\maketitle

\thispagestyle{empty}

\begin{abstract}
Repeated occurrences of serial firing sequences of a group of neurons
with fixed time delays between neurons are observed in many
experiments involving simultaneous recordings from multiple
neurons. Such temporal patterns are potentially indicative of
underlying microcircuits and it is important to know when a repeatedly
occurring pattern is statistically significant. These sequences are
typically identified through correlation counts, such as in the
two-tape algorithm of Abeles and Gerstein \cite{Abeles1988}. In this paper we present
a method for deciding on the significance of such correlations by
characterizing the influence of one neuron on another in terms of
conditional probabilities and specifying our null hypothesis in terms
of a bound on the conditional probabilities. This method of testing
significance of correlation counts is more general than the currently
available methods since under our null hypothesis we do not assume
that the spiking processes of different neurons are independent. The
structure of our null hypothesis also allows us to rank order the
detected patterns in terms of the strength of interaction among the
neurons constituting the pattern. We demonstrate our method of
assessing significance on simulated spike trains involving
inhomogeneous Poisson processes with strong interactions, where the
correlation counts are obtained using the two-tape
algorithm \cite{Abeles1988}.
\end{abstract}

\section{Introduction}

Detection of temporal firing patterns among groups of neurons is an important task in neuroscience as such patterns may be indicative of functional cell assemblies or microcircuits present in the underlying neural tissue \cite{Gerstein1985, Palm1988}. Analysis methods such as the two-tape algorithm of Abeles and Gerstein \cite{Abeles1988,Abeles2001} have been developed to discover repeating occurrences of precise firing sequences in simultaneous recordings from multiple neurons. The two-tape algorithm has been used to assess precisely timed activity patterns {\em in vivo} \cite{Nadasdy1999}, in slices {\em in vitro} \cite{Ikegaya2004}, and recently in cultures of dissociated cortical neurons \cite{Rolston2007}. The two-tape algorithm, as well as alternative methods for identifying spike coincidences such as the unitary event detection algorithm of Grun \cite{Grun2002}, determines statistical significance of the discovered patterns under a null hypothesis of independence among the neurons \cite{Brown2004}. In this paper, we present a significance test that allows ``weak interactions'' to be included in the null hypothesis by characterizing the strength of influence among neurons in terms of a conditional probability, which is a very natural way to think about synaptic interactions between neurons). 

Let us write a 3-neuron pattern as 
$A\stackrel{T_1}{\rightarrow}B\stackrel{T_2}{\rightarrow}C$, denoting a firing sequence
where $A$ is followed by $B$ after a delay of $T_1$ and $B$ is followed by $C$ after a delay of
$T_2$. Most methods 
count the occurrences of this pattern by essentially finding correlations among 
time shifted spike trains from $A$, $B$, and $C$. There are also many methods to determine the statistical significance
of these patterns based on how many times they occur 
(see \cite{Gerstein2004} and references therein). To assess whether a given number of repetitions of the 
pattern is significant, one generally employs a null hypothesis that 
 assumes that all neurons spike as (possibly inhomogeneous) Poisson processes 
and that different neurons are independent. Based on this one can analytically 
calculate a bound on the number of repetitions required to make a pattern significant 
 in the sense of being able to reject the null hypothesis at a 
 given level of confidence. There are also methods to assess significance through empirical means 
\cite{Bienenstock2003}. 
In these methods, one generates many surrogate data streams of spike trains by
systematically perturbing the spikes in the original data and then assesses significance
of a pattern by noting the difference in counts (or in any other statistic derived from 
such correlation counts)  for these patterns in the original
data and in the surrogate streams. In these ``jitter'' methods, often the 
implicit null hypothesis is also independence.  

When a sequential firing pattern like $A\stackrel{T_1}{\rightarrow}B\stackrel{T_2}{\rightarrow}C$
is declared as significant by any of these methods, 
the underlying idea is that we can conclude that $A$, $B$, and $C$ influence 
each other in a sequential fashion because we are able to reject the 
null hypothesis of independence. However, these methods do not say anything about the
 ``strength of influence'' among neurons $A$, $B$, and $C$. We present a method for analyzing 
statistical significance of sequential firing patterns that also allows rank ordering of 
significant patterns in terms of the strength of influence among participating neurons. Thus our method extends the 
currently available techniques of significance analysis. 

We represent the strength of influence of $A$ on $B$ by the conditional 
probability that $B$ would fire after the delay $T_1$ given that $A$ has 
fired now, which we denote as $e(B|A,T_1)$. Our null hypothesis is then stated in terms of a bound on all such 
pairwise conditional probabilities and we develop a statistical significance test for 
rejecting the null hypothesis. By changing the parameter of the null hypothesis (which is the bound on the conditional probability), we are able to see which patterns are significant at different levels of strength of influence. 

Another interesting consequence of our approach is that since the null hypothesis is stated in terms of bounds on pairwise conditional probabilities, our null hypothesis includes 
not only all models of independent neurons but also some models of dependent or 
interacting neurons. This is more general than what is available in 
current methods. Intuitively, declaring a 
pattern like $A\stackrel{T_1}{\rightarrow}B\stackrel{T_2}{\rightarrow}C$ 
as significant should mean that we can conclude that there are
``strong'' causative connections from $A$ to $B$ and from $B$ to $C$ with the 
indicated delays. Hence rejecting a null hypothesis that includes not only 
independent neuron models but also models of neurons that are ``weakly'' 
dependent is more appropriate. Our method specifies 
these ``weak interactions'' in terms of bounds on conditional probabilities. 

The rest of the paper is organized as follows. In Section~\ref{sec:meth} we present  
our significance test. We first explain our composite null hypothesis and then 
develop a test of significance. In Section~\ref{sec:results} we demonstrate the 
effectiveness of the method through computer simulations. Spike trains are generated by a network of neurons modeled as interdependent inhomogeneous Poisson processes. We show that our method rank orders significant 
patterns. Surprisingly, it is also quite effective in situations where some of the 
assumptions of our theoretical analysis are not valid. We conclude the paper 
in Section~\ref{sec:diss} with potential extensions and a discussion of strengths and weaknesses. 

\section{Methods}
\label{sec:meth}
\begin{flushleft} {\bf Correlation Count} \end{flushleft}
For simplicity of exposition, we first explain the method for a sequential firing 
pattern of only two neurons. Consider a pattern $A\stackrel{T}{\rightarrow}B$. 
Suppose we find the number of repetitions of this pattern in the data using 
simple correlation as follows. Let $t_1, t_2, \ldots t_n$ denote all time 
instants at which there is a spike from any neuron in the data. Let
\begin{equation}
f_{AB} = \sum_{i=1}^n \: I_{A}(t_i) I_{B}(t_i + T)
\label{eq:2-correlation}
\end{equation}
where for any neuron $x$, $I_{x}(t) = 1$ if there is a spike from $x$ at time $t$ and 
zero otherwise. Note that $f_{AB}$ is simply a correlation integral which counts the number of spikes from $A$ that are followed by $B$ with a delay of exactly 
$T$ units, and hence counts the number of repetitions of our pattern. 
If we want to allow for some small random variations in the delay we can define 
the indicator variable $I_{x}(t)$ to take value 1 if there is a spike in a time interval 
of length $\Delta T$ centered around $t$. (For example, we can take $\Delta T$ to be 
the time resolution in our measurements). From now on we assume that delays are 
always over some such small intervals.  

There are many methods to calculate correlation counts, for example the two-tape 
algorithm of Gerstein and Abeles \cite{Abeles1988} and some of its recent variations \cite{Tetko2001}.
Most current methods for detecting serial firing patterns rely on such correlations.  
Since the focus of this paper is on statistical significance (and not on 
computational efficiency), we simply assume that 
one can calculate such counts for pairs of neurons and for various delays $T$ of interest. The question then, is ``how large should the count be to conclude that the pattern represents a ``strong'' influence of $A$ on $B$?''

Since we want to address this question in a classical hypothesis testing framework, 
we need to choose a null hypothesis that includes as many models as possible 
of interdependent neurons without any ``strong'' influences between pairs of neurons. 
Then, if we can calculate (or bound) the probability under the null hypothesis 
that $f_{AB}$ is above a threshold, we get a test of statistical significance. 
As stated earlier, we want the null hypothesis to contain a parameter to denote the 
strength of influence so that we can rank order all significant patterns. 

\begin{flushleft} {\bf Strength of influence as conditional probability}\footnote{See \cite{sastrymanu} for the original formalism of using conditional probability as a measure of interaction strength and deriving bounds on counts}\end{flushleft}
We propose that the strength of influence between any pair of neurons can be 
characterized in terms of a conditional probability. Let $e(B|A,T)$ denote the 
conditional probability that $B$ will fire at time $T$ (or more precisely, in 
a time interval $[T - \frac{\Delta T}{2}, \  T+\frac{\Delta T}{2}]$) given that $A$ has fired at time 
zero. If, for example, there is a strong excitatory connection from $A$ to $B$, this 
probability would be large. If, on the other hand, $A$ and $B$ are independent then 
$e(B|A,T)$ would be the same as the unconditional probability of $B$ firing in an 
interval of length $\Delta T$. (For example, if we take that $\Delta T = 1 ms$ and the average firing rate of $B$ is $20 Hz$, then this unconditional probability would be 
about 0.02). We note here that this conditional probability is well defined even if the two neurons are not directly connected through a single synapse. If the pair is directly connected, then $T$ can be a typical mono-synaptic delay; otherwise $T$ can be a multiple of the mono-synaptic delay. In either case, our task is to find whether a pattern with a specific value for $T$ is significant. 

This conditional probability is a good scale on which to say whether 
the influence of $A$ on $B$ is ``strong''. Our main assumption here is that this 
conditional probability is not time dependent. That is, the probability that 
$B$ fires in an interval $[t+T-\frac{\Delta T}{2}, \ t+T+\frac{\Delta T}{2}]$ given that $A$ has 
fired at $t$ is the same for all $t$ for the time period of observations that we are 
analyzing. Some recent analysis of spike trains from neural cultures \cite{Feber2007} suggests that such an assumption is justified. Note that this assumption does not require the firing rates of neurons to not be time-varying. As a matter of fact, one of the main mechanisms by which this conditional probability is realized 
is by having a spike from $A$ affect the rate of firing by $B$ for a short duration 
of time. Thus, the neurons would have time-varying firing rates even when the 
conditional probability is not time-varying. Our assumption is only that every time $A$ spikes, it has the same chance of 
eliciting a spike from $B$ after a delay of $T$, i.e. there are no appreciable changes in synaptic efficacies during the period in which the data is gathered. 
 
When analyzing the significance of repeating serial patterns, what we are interested in is hypothesizing causative chains. Hence the strength of a pattern should be related to the propensity that a spike from $A$ has on eliciting a spike from $B$, which can be conveniently represented by the conditional probability of $B$ spiking given that $A$ has spiked. In all serial firing patterns of interest, the constancy of delays in multiple repititions are important. Hence we defined the conditional probability with respect to a specified delay. Capturing influences among neurons through this conditional probability allows us to formulate an interesting compound null hypothesis in terms of a bound on these probabilities. 

\begin{flushleft}{\bf Compound null hypothesis}\end{flushleft}
Now we propose the following {\em compound} null hypothesis. Any model of interacting neurons 
is in our null hypothesis if it satisfies: $e(Y|X,T) \leq e_0$ for all neurons $X,Y$ 
and a set of specified delays $T$, where $e_0$ is a user-chosen constant.   
The exact mechanism by which a spike output by $A$ affects the spiking of $B$ is 
immaterial here. Whatever be the mechanism, if the resulting conditional probability 
is less than $e_0$ then that model would be included. Thus our compound null 
hypothesis includes many models of interdependent neurons where whether or not a 
neuron spikes can depend on actual spikes of other neurons (unless 
we choose $e_0$ to be very small). The idea is that we choose an $e_0$ based on how strong we want 
the influences to be before we agree to say there is a causative influence of $A$ 
on $B$. For example, as given earlier, with $\Delta T = 1 ms$, and an average rate of 
firing of $20 Hz$, the conditional probability is 0.02 if the neurons are independent. 
So, if we choose $e_0=0.4$, it means that we agree to call an influence strong if 
the conditional probability is twenty times what you would see if the neurons were independent. More importantly, if we 
have a test of significance for this null hypothesis, then by varying $e_0$ we can rank order different significant patterns in terms of the strength of influence.
 
\begin{flushleft}{\bf Significance Test}\end{flushleft}
To get a test for statistical significance we need to calculate a bound on the
probability that, under this null hypothesis, the count $f_{AB}$ is 
above a given threshold. For this, consider the following stochastic model. Suppose 
$L$ is the total time duration of the data and let the random variable $N_A(L)$ denote the total number of 
spikes by neuron $A$ during this time. Define 
\begin{equation}
S_{AB} = \sum_{i=1}^{N_A(L)} \: X_i
\label{eq:S}
\end{equation}
where $X_i$ are independent and identically distributed 0-1 random variables with 
\begin{equation}
\mbox{Prob}[X_i = 1] = p \ \ \ ( = 1 - \mbox{Prob}[X_i=0]).
\label{eq:xis}
\end{equation}
If we take $p=e(B|A,T)$, it is easy to see that $S_{AB}$ is a random variable equivalent to $f_{AB}$ since every time there is a spike from $A$, with probability $p$ a spike from $B$ would follow with the appropriate delay. Also, per our assumption,  
every time $A$ spikes a spike from $B$ with the appropriate delay occurs with the same probability regardless of $B$'s spike history. This implies that the $X_i$ in the definition of $S_{AB}$ can be assumed to be {\em independent and identically distributed}. 

Now we assume that the spiking of $A$ is Poisson. (Note that we are {\em not} assuming that 
spiking by $B$ is Poisson and, more importantly, as per our null model the spiking process 
of $B$ is not independent of that of $A$.)\footnote{Indeed when spiking by $A$ is Poisson, the spiking of $B$ would not be Poisson if there is sufficient influence of $A$ on $B$. Also, in Section \ref{sec:results} we show empirically that our hypothesis testing method is effective even in cases when $A$ is not Poisson.} 
Then the random variable $S_{AB}$ is such that we are accumulating a random variable $X_i$ every time an event of a Poisson process happens. This implies that, since the $X_i$ are 0-1 random variables, $S_{AB}$ is also a 
Poisson random variable \cite[Ch.2.5]{Ross1996}.\footnote{An equivalent 
way of looking at this is to consider the Poisson process of spikes from $A$ and suppose that we classify 
each spike as type-Y with probability $p$ and as type-N with probability $(1-p)$. Then  
it can be shown that the sequence of type-Y spikes (and also the type-N spikes) 
would constitute a Poisson process. Here, the classification of each $A$ spike is dependent on whether or not there is a spike from $B$ after the appropriate delay and is independent of everything else.} The mean and variance of $S_{AB}$ are given by \cite[Ch.2.5]{Ross1996}
\begin{eqnarray}
E S_{AB} & = &  E[N_A(L)] \: E[X_i] \nonumber \\
\mbox{Var}(S_{AB}) & = & E[N_A(L)] E[X_i^2] 
\end{eqnarray}
Let the rate of the Poisson process for $A$ be $\lambda_A$. Then, $E[N_A(L)] = L \lambda_A$. 
Since, $E[X_i] = E[X_i^2] = p$, $S_{AB}$ is a Poisson random variable with 
expectation (and variance) $\lambda_S = p L \lambda_A$.

\begin{flushleft}{\bf Computing the threshold}\end{flushleft}
Let $Z$ be a Poisson random variable with mean $\lambda_Z$. Suppose the allowed Type~I 
error in our hypothesis test is $\alpha$. Let $M$ be the smallest number satisfying 
\begin{equation}
\mbox{Prob}[ Z > M] \leq \alpha.
\label{eq:test}
\end{equation}
Given $\alpha$ and $\lambda_Z$, we can calculate the $M$ needed to satisfy the 
above using the Poisson distribution. For a Poisson random variable, the probability 
on the LHS of Eq.~(\ref{eq:test}) is monotonically increasing with $\lambda_Z$ 
as long as $\lambda_Z < M$.  We know that $S_{AB}$ is Poisson with mean 
$e(B|A,T) L \lambda_A$. Under our null hypothesis, we have $e(B|A,T) < e_0$. 
Hence, if we take $\lambda_Z = e_0 L \lambda_A$ and calculate the $M$ needed to 
satisfy Eq.~(\ref{eq:test}), then we have, under our null hypothesis, 
\begin{equation}
\mbox{Prob}[ S_{AB} > M] \leq \alpha.
\label{eq:test-fn}
\end{equation}
Since, as discussed earlier, the random variable $S_{AB}$ represents the count $f_{AB}$, 
the above $M$ is the threshold on the count to reject our null hypothesis and 
hence conclude that the pattern found is significant. 

The test of statistical significance is as follows. {\em Let $e_0$ be the 
bound on conditional probability that we chose for our null hypothesis. Let 
$\alpha$ be the allowed Type~I error. Let $\lambda_A$ be the rate of firing for 
the first neuron in the pattern. Set $\lambda_Z = e_0 L \lambda_A$. Using the 
cumulative distribution of a Poisson random variable with parameter $\lambda_Z$, we 
calculate the $M$ needed to satisfy Eq.~(\ref{eq:test}). This $M$ is the threshold on 
the count of the pattern for us to be able to reject the null hypothesis and 
declare the pattern to be significant.}

\begin{figure}[htbp]
\centering
\includegraphics[scale=0.4,trim=0in 0in 0in 0in, clip]{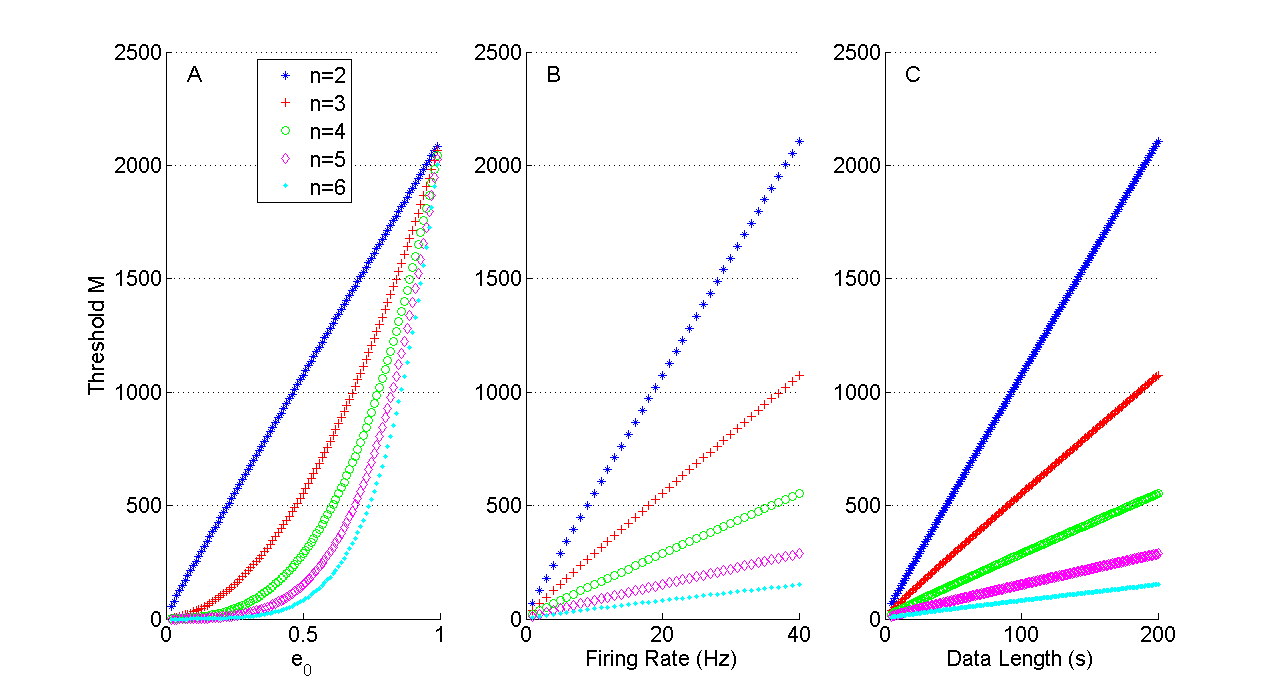}
\caption{Thresholds calculated per Eq.~(\ref{eq:test-fn}) for $n$=2,3,4,5 and 6-neuron patterns with $\alpha$=0.01. {\em A:} Threshold as a function of pattern strength for $\lambda_A = 20 Hz$ and $L = 100 s$. {\em B:} Threshold as a function of firing rate for $e_0=0.5$ and $L = 100 s$. {\em C:} Threshold as a function of data length for $\lambda_A = 20 Hz$ and $e_0 = 0.5$. We can see that the threshold is a smooth function of all parameters. The shape of the threshold curves are similar for other values of Type~I error.}
\label{fig:thresh}
\end{figure}

To calculate this threshold, we need $\lambda_A$. This is easily estimated from the 
data as the average rate of firing for neuron $A$. A simple parametric study shows that the threshold $M$ is well-behaved (see Fig. \ref{fig:thresh}).

\begin{flushleft}{\bf Extending to longer patterns}\end{flushleft}
Here we explain how the signficance test can be extended to patterns involving more than two neurons. Suppose we are considering the pattern $A\stackrel{T_1}{\rightarrow}B\stackrel{T_2}{\rightarrow}C$. We assume we 
get the count of $f_{ABC}$ by taking 3-point correlations.\footnote{In general, taking $n$-point correlations like this for all possible $n$-tuples of patterns is computationally expensive. For this reason, such correlation counts are obtained only for patterns of length 3 or 4 in most cases. Here we are only explaining how our test can be extended to assess significance of longer patterns provided we can get such correlation counts.}  
As before we define $S_{ABC}$ as a sum of 0-1 random variables $X_i$. Now, we want $X_i$ to be 1 if 
the spike by $A$ is first followed by $B$ and then by $C$ with the indicated delays. 
Hence we take $p$ to be $e(B|A,T_1) e(C|B,T_2) L \lambda_A$.\footnote{This 
calculation for $p$ is correct if all influence of $A$ on $C$ comes only through $B$. This is a reasonable assumption if we want significant patterns to represent a chain of triggering events. In such a case, $S_{ABC}$ would be same as $f_{ABC}$. Even if there are other paths for $A$ to influence $C$, this value of $p$ would represent a lower bound on the probability of the pattern occurring at any spike from $A$. Hence, with this $p$, $S_{ABC}$ would be less than $f_{ABC}$ and hence the threshold on the count calculated using this $S_{ABC}$ would be sufficient in a hypothesis testing framework.}  
Under the null hypothesis, each of these conditional probabilities are less than $e_0$. Hence, if we calculate the $M$ as needed in Eq.~(\ref{eq:test}) with $\lambda_Z = (e_0)^2 L \lambda_A$, then we get the needed threshold on the counts 
for this pattern. Now the method can easily be generalized to a pattern involving $n$ neurons. Let the 
first neuron in the pattern be $A$. Then we calculate the threshold $M$ needed using 
Eq.~(\ref{eq:test}) with $\lambda_Z = (e_0)^{n-1} L \lambda_A$. 

The main point of the above analysis is that if the first neuron in a chain is Poisson then the counts of the pattern
would be Poisson even if the other neurons in the chain are not Poisson. As a matter of fact, when the relevant conditional probabilities are high, the other neurons would not be Poisson. It was observed earlier \cite{Abeles2001} that the correlation counts can be assumed to be Poisson even when the neurons have time-varying rates and hence are not strictly Poisson (but may be inhomogenous Poisson). The analysis presented here can be viewed as a theoretical justification for this observation. 

\section{Results}
\label{sec:results}
\begin{figure}[htbp]
\centering
\includegraphics[scale=1,trim=0in 0in 0in 0in, clip]{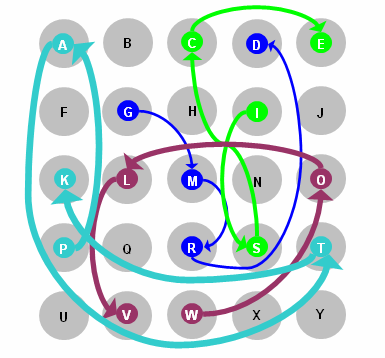}
\caption{Network of 25 neurons with 4-neuron chains of various connection strengths (in terms of conditional probabilities) and synaptic delays (in $ms$). The strengths and delays in the chains are as follows (denoted Neuron 1[delay, strength]-Neuron 2): 
G[2,0.2]-M[3,0.2]-R[4,0.2]-D, I[5,0.4]-S[4,0.4]-C[3,0.4]-E, W[3,0.6]-O[5,0.6]-L[2,0.6]-V, and P[4,0.8]-A[2,0.8]-T[5,0.8]-K.  All neurons in the network have a background firing rate of $20Hz$ and a refractory period of $1 ms$. In simulations we update the firing rate of each neuron every $\Delta T = 1 ms$.}
\label{fig:network}
\end{figure}

\begin{flushleft} {\bf Spike train simulator}\end{flushleft}  
In this section, we present results from computer simulations to demonstrate the effectiveness of our method. We used a simulator for generating spike data from a network of 25 interconnected neurons labeled $A$ through $Y$ as shown in Fig. \ref{fig:network}. There were four chains, each four neurons in length, with strong connections in the network (G-M-R-D, I-S-C-E, W-O-L-V, and P-A-T-K). The spiking of each neuron was an inhomogeneous Poisson process whose background firing rate of $20 Hz$ was modulated at time intervals of $\Delta T = 1 ms$ based on input received from neurons to which it was connected. Each synapse interconnecting neurons was characterized by a delay which was an integral multiple of $\Delta T$ and a strength in terms of a conditional probability. Using the notation from Section \ref{sec:meth}, the connections in the four chains were specified as follows (delays in $ms$):

\begin{eqnarray} \nonumber
e(M|G,2)=e(R|M,3)=e(D|R,2)=0.2\\ \nonumber
e(S|I,5)=e(C|S,4)=e(E|C,3)=0.4\\ \nonumber
e(O|W,3)=e(L|O,5)=e(W|L,2)=0.6\\ \nonumber
e(A|P,4)=e(T|A,2)=e(K|T,5)=0.8\\ \nonumber
\end{eqnarray}
So among the four chains, the conditional probabilites ranged from 0.2 to 0.8 (with G-M-R-D being the weakest and P-A-T-K being the strongest) and the synaptic delays were between 2 and 5 $ms$. In some simulations the network also had many other interconnections in addition to the chains, where we connected each neuron to 25\% of all the other neurons randomly. As stated earlier, with a $20 Hz$ average firing rate the unconditional probability of a neuron firing in $\Delta T$ is about 0.02. Hence we chose the strength (again in terms of conditional probabilities) of these random connections as uniformly distributed over the range $[0.01, \ 0.04]$, which is a factor of 2 on either side of the case of independence. {\em Note that this means that the firing rate of a neuron (under the influence of random synapses) varies over the range of $\lambda/2$ to $2 \lambda$ where $\lambda$ is the nominal background firing rate}. The synaptic delays for the random connections were uniformly distributed between 2 and 5 $ms$. All neurons also had a refractory period of $1 ms$. Further details on the simulator can be found in Appendix B. 

\begin{figure}[htbp]
\centering
\includegraphics[scale=0.4,trim=0in 0in 0in 0in, clip]{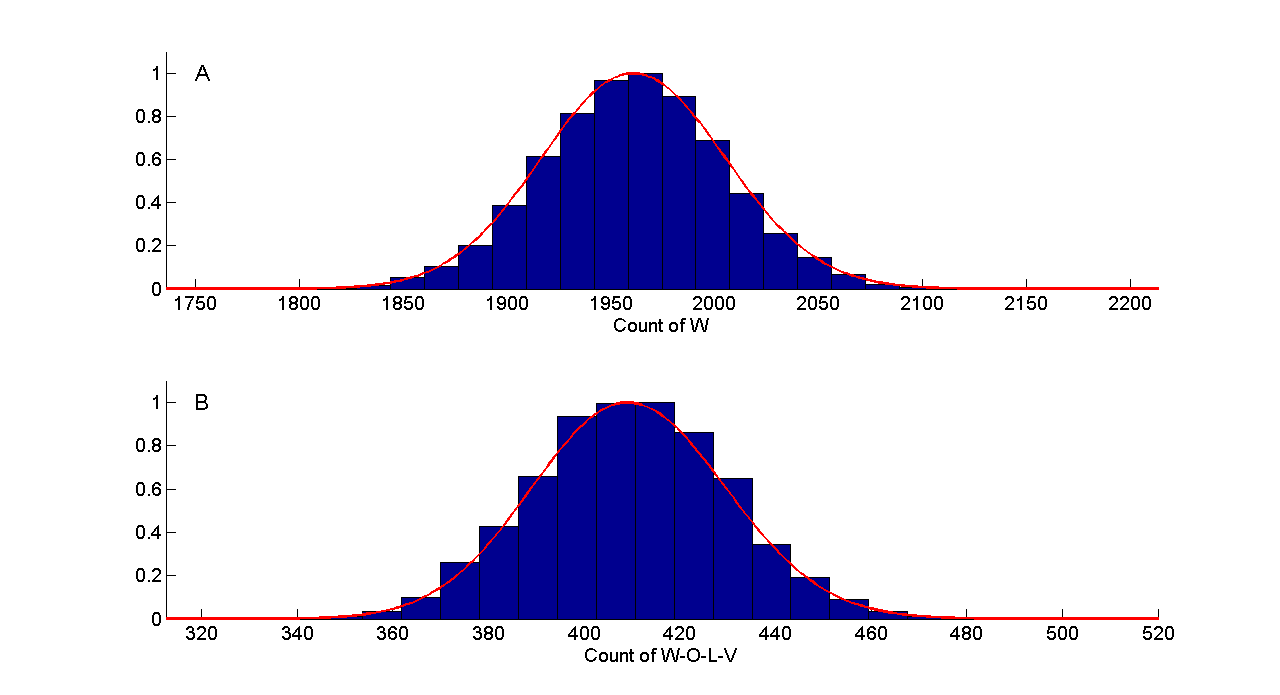}
\caption{First neuron and 4-neuron pattern count histograms of chain with strength 0.6 (W-O-L-V) in a 25-neuron network {\em with no random connections}, $\lambda_A = 20 Hz$, $L =100 s$, refractory period $1 ms$, 5,000 replications. {\em A:} Red curve is the Poisson distribution with $\lambda$ equal to the mean count of $W$ in the samples. The count of $W$ appears to be approximately Poisson. {\em B:} Red curve is the Poisson distribution with $\lambda$ equal to the mean count of W-O-L-V in the samples. The count of W-O-L-V appears to be approximately Poisson.}
\label{fig:charpois}
\end{figure}

\begin{figure}[htbp]
\centering
\includegraphics[scale=0.4,trim=0in 0in 0in 0in, clip]{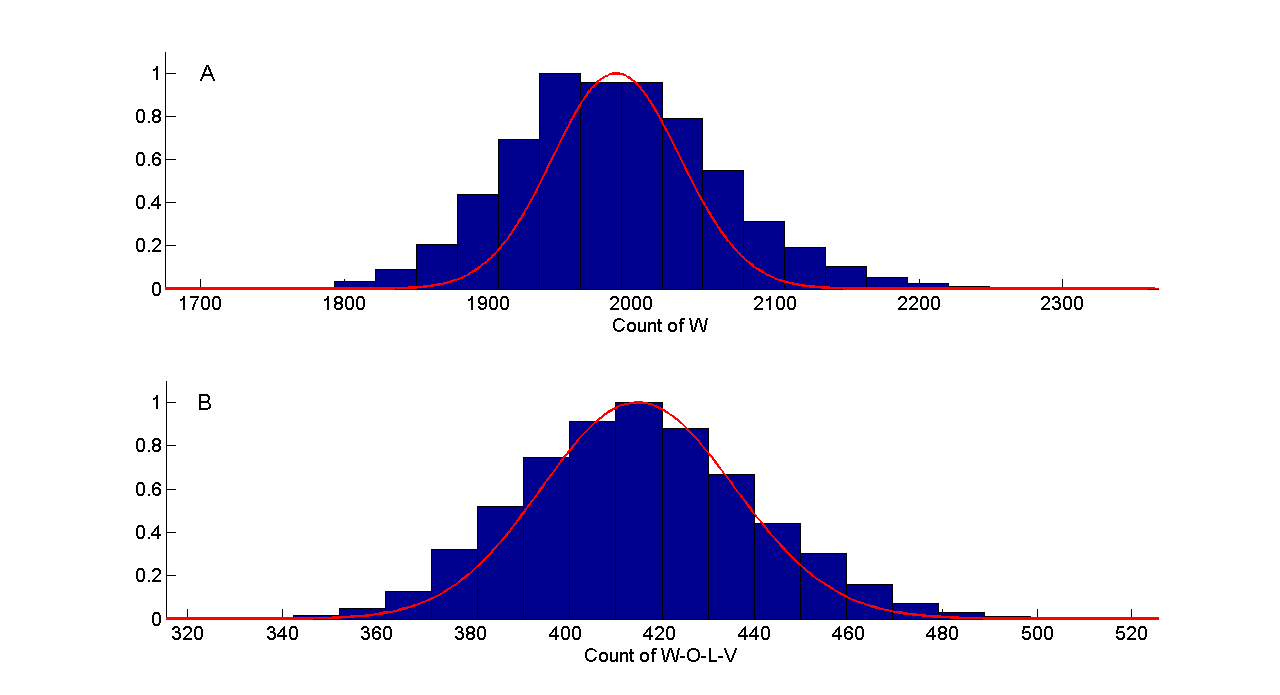}
\caption{First neuron and 4-neuron pattern count histograms of chain with strength 0.6 (W-O-L-V) in a 25-neuron network {\em with random connections}, $\lambda_A = 20 Hz$, $L = 100 s$, refractory period $1 ms$, 5,000 replications. {\em A:} Red curve is the Poisson distribution with $\lambda$ equal to the mean count of $W$ in the samples. The count of $W$ deviates from Poisson. {\em B:} Red curve is the Poisson distribution with $\lambda$ equal to the mean count of W-O-L-V in the samples. The count of W-O-L-V deviates from Poisson.}
\label{fig:charnonpois}
\end{figure}

\begin{flushleft}{\bf First neuron and pattern characteristics}\end{flushleft}
We conducted simulations that verified the result derived in Section \ref{sec:meth} that the occurrences of a pattern are Poisson-distributed when the first neuron in the pattern spikes according to a Poisson process. To do this we simulated the 25-neuron network shown in Fig. \ref{fig:network} for 100 seconds without any random connections. Since the only connections among neurons were the four chains (G-M-R-D, I-S-C-E, W-O-L-V, P-A-T-K), the spiking of the first neuron in each pattern ($G$, $I$, $W$, $P$) should approximately follow a Poisson process. (Even here the spiking of the first neurons are not exactly Poisson-distributed since they have a refractory period of 1 ms. However, since this refractory period is small compared to the interspike intervals when firing at $20 Hz$, the deviations from the Poisson firing due to the refractory period are small). We repeated the simulation 5,000 times obtaining the counts of $W$ and the counts of the pattern W-O-L-V. In Fig. \ref{fig:charpois} we show the histogram of these counts compared to a Poisson distribution with the rate parameter set as the sample mean for each count. The variance-to-mean ratio, or the Fano factor, for the count of $W$ is 0.97. Since the Fano factor of a Poisson random variable is 1 this indicated that the first neuron was approximately Poisson, and indeed a $\chi^2$ goodness-of-fit test failed to reject a null hypothesis of Poisson (P-value>0.05, test performed following the guidelines in \cite{Nair1994}). The theory developed in Section \ref{sec:meth} tells us that when the first neuron in a pattern is approximately Poisson the count of the pattern will also be approximately Poisson, and our empirical results confirmed this (W-O-L-V Fano factor 1.01, $\chi^2$ goodness-of-fit test P-value>0.05). Then we repeated the simulations this time allowing random connections between all neurons with 25\% connectivity. We can see in Fig. \ref{fig:charnonpois} that when there are random connections in the network the first neuron in the pattern ($W$) and the pattern itself (W-O-L-V) are no longer approximately Poisson-distributed (Fano factor 2.66 and 1.51 for $W$ and W-O-L-V respectively). The $\chi^2$ goodness-of-fit test rejects the null hypothesis that these counts are Poisson ($W$: P-value<0.01, W-O-L-V: P-value<0.01). Although the theory behind our significance thresholds assumes the first neuron spikes according to a Poisson process, we will now demonstrate empirically that even when the first neuron is not Poisson-distributed (due to random connections present in the network) we are still able to rank order the relative strength of patterns effectively using our significance thresholds.\\ 

\begin{figure}[htbp]
\centering
\includegraphics[scale=0.4,trim=0in 0in 0in 0in, clip]{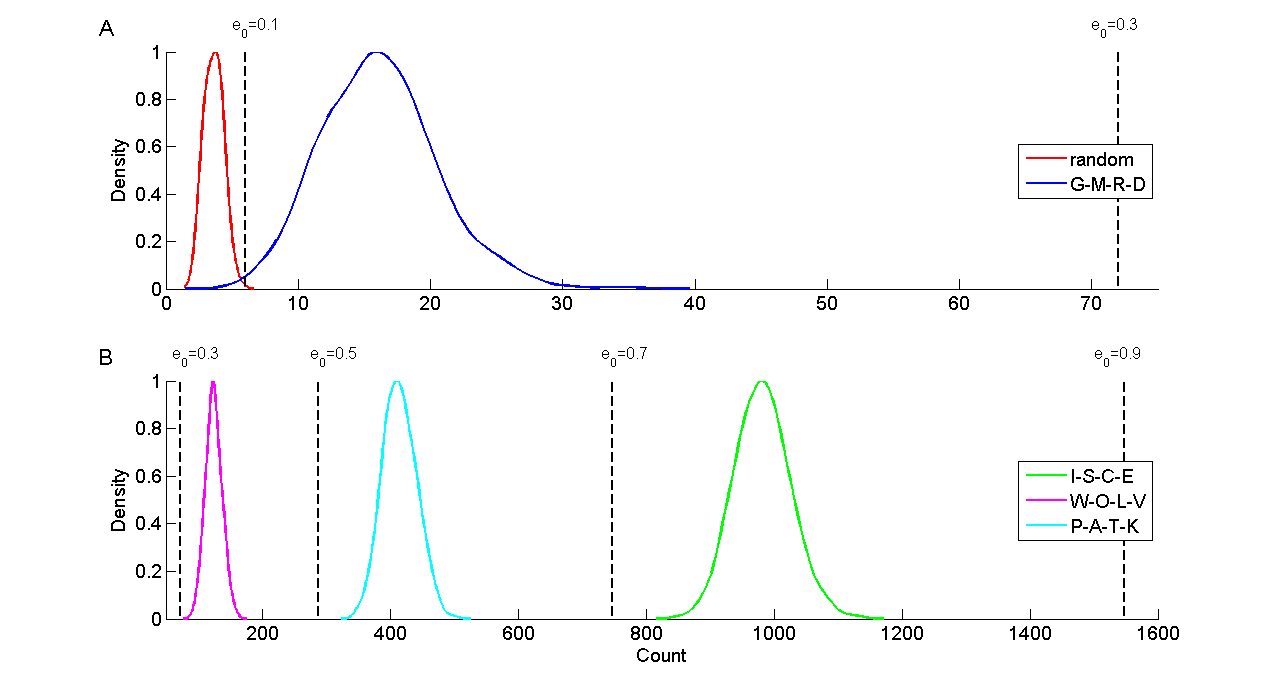}
\caption{Rank ordering of different patterns in a 25-neuron network, $L = 100 s$, 1,000 replications. {\em A:} maximum random 4-neuron pattern counts (not involving any neurons in the four chains); pattern G-M-R-D counts (strength 0.2); threshold for $e_0=0.1$. {\em B:} I-S-C-E (strength 0.4), W-O-L-V (strength 0.6), and P-A-T-K (strength 0.8) pattern counts; thresholds for $e_0=0.3,0.5,0.7,0.9$.}
\label{fig:rankorder}
\end{figure}

\begin{flushleft}{\bf Rank ordering of signficant patterns}\end{flushleft}  
Even when the pattern counts are not Poisson-distributed (as for the simulations shown in Fig. \ref{fig:charnonpois}), we are still able to rank order the relative strengths of the patterns as shown in Fig. \ref{fig:rankorder}.  We also see that the line showing the threshold count corresponding to $e_0=0.1$ is able to distinguish the counts of pattern G-M-R-D from the ``maximum'' of random 4-neuron pattern counts not involving any of the neurons in the four chains. 

\begin{figure}[htbp]
\centering
\includegraphics[scale=0.4,trim=0in 0in 0in 0in, clip]{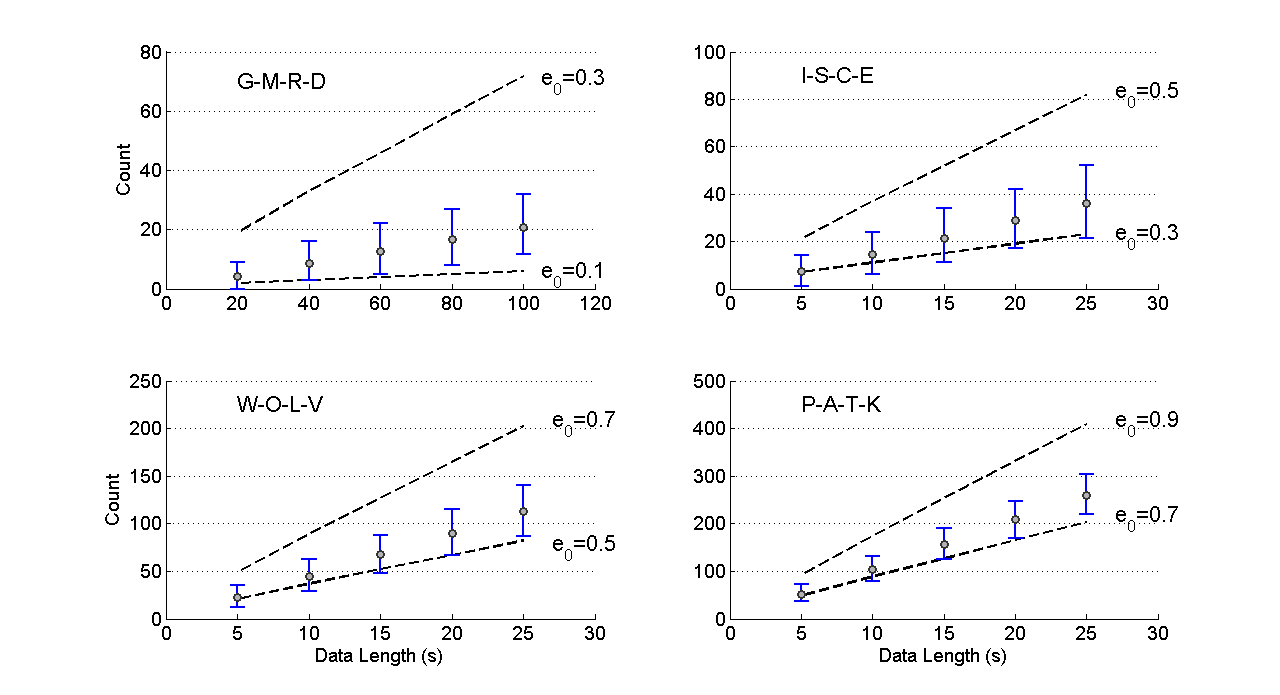}
\caption{Data length sufficiency for the chains of various strengths in the 25-neuron network. Open circle is mean count, error bars are $1^{st}$ and $99^{th}$ percentile out of 1,000 replications.}
\label{fig:datasuff}
\end{figure}

\begin{flushleft}{\bf Data requirements}\end{flushleft}  
To determine how much data is required to rank order we simulated the network with random connections to all neurons (again with 25\% connectivity) for various lengths of time, and for each data length compared the counts of the patterns to the thresholds corresponding to $e_0$ values which are 0.1 greater than and 0.1 less than the known connection strength of the chain producing the pattern. We found that the amount of data needed to achieve the desired resolution depends on the chain strength. We can see from Fig. \ref{fig:datasuff} that for pattern G-M-R-D (strength 0.2) around 60 seconds of data is sufficient, while for patterns I-S-C-E, W-O-L-V, and P-A-T-K (strengths 0.4, 0.6, and 0.8) we need as little as 15 to 25 seconds of data. The data requirements are also dependent on firing rate, and with $\lambda = 5 Hz$ we find that 300 seconds of data is sufficient for the weakest pattern (G-M-R-D). We then used this firing rate and data length to demonstrate how our techniques can enhance the significance analysis of counts obtained using the two-tape algorithm of \cite{Abeles1988}.

\begin{figure}
	\centering
		\includegraphics[scale=0.4,trim=0in 0in 0in 0in, clip]{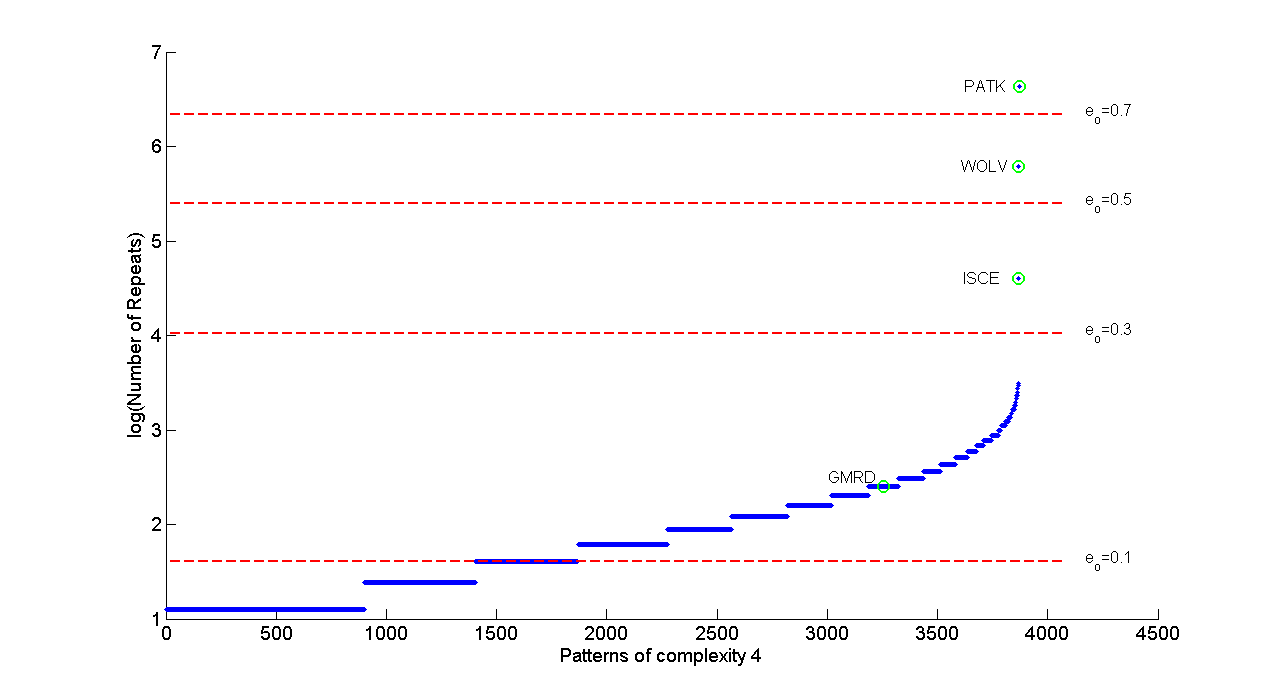}
	\caption{Results of mining a 25-neuron network simulation (baseline firing frequency $5 Hz$, $L = 300 s$) with the two-tape algorithm. There were 3,870 4-neuron patterns that repeated more than twice. We plot the logarithm of how often each pattern repeated, and also plot our threshold for statistical significance of 4-neuron patterns for various $e_0$.}
	\label{fig:mining4node}
\end{figure}

\begin{flushleft}{\bf Enhanced significance analysis of two-tape algorithm counts}\end{flushleft}
Abeles and Gerstein \cite{Abeles1988} provide a formula for calculating the expected number of patterns of a particular description that will occur $r$ times in data of length $L$ if the neurons spike according to independent Poisson processes. Based on this formula, if we had 25 independent spike trains and 300 seconds of data we would expect to find 3.86 patterns of complexity (length) four that repeat at least twice (with $\lambda = 5 Hz$, $\Delta T = 1 ms$, and the total time span between the spikes of the first and fourth neuron constrained to be $15 ms$ or less). When we ran a simulation with these parameters and then mined the data with the two-tape algorithm we found six 4-neuron patterns that repeated twice or more and satisfied the temporal constraint. Again following \cite{Abeles1988}, we calculated that this is not a statistically significant excess of patterns for a Poisson random variable with a mean of 3.86 (P-value>0.05). However if it had been, since it is a small number of patterns each pattern could be investigated further by the experimenter to determine which particular patterns out of the six are actually of interest. On the other hand, if the number of patterns found is large (and statistically significant) it is not practical to designate all patterns of that description as candidates for further investigation. To illustrate this, we repeated the simulation but instead of having independent neurons we had chains of connected neurons as shown in Fig. \ref{fig:network}, as well as random connections between all neurons with 25\% connectivity as described previously. This time when we mined with the two-tape algorithm we found 3,870 4-neuron patterns that repeated at least twice. Here we need some additional criteria to select which of these individual patterns are most likely to be significant and the best candidates for further analysis. Abeles and Gerstein \cite{Abeles1988} remarked that this selection process is very important, and called for future research to be conducted in this area to devise selection methods beyond their suggested strategy of repeating their analysis procedure for different subgroupings of the patterns (with the hopes of finding a smaller group of patterns that is significant which can then be investigated further to find the individual patterns responsible). Our framework of a compound null hypothesis based on conditional probabilities can be very useful as the selection criteria. By having different values for $e_0$ in the null hypothesis, we can ask what patterns are significant at what value of $e_0$ and thus rank order patterns according to their relative strength. We demonstrate this in Fig. \ref{fig:mining4node} as our thresholds at various $e_0$ are able to separate the three strongest patterns (I-S-C-E, W-O-L-V, and P-A-T-K) from the rest based on the counts obtained using the two-tape algorithm. We also see that when mining with a threshold of $e_0=0.1$ there are a considerable number of false positives (i.e. there are other 4-neuron patterns that are more frequent than G-M-R-D, our weakest pattern). This is due to subpatterns (e.g. P-A-T) of the stronger patterns being very frequent, and as a consequence 4-neuron patterns such as P-A-T-X occur frequently even if there is no strong connection between T and X. Future work will address this issue. 

\section{Discussion}
\label{sec:diss}


In this paper we proposed a method of assessing significance of serial firing patterns using 
correlation counts as the statistic. There are two attractive features of this method.  
First, we can rank order significant patterns in terms of their relative ``strength''. For this 
we represent the strength of influence of $A$ on $B$ by the conditional probability 
that $B$ fires after a prescribed delay following $A$. We state our composite null 
hypothesis in terms of a parameter $e_0$ which is an upper bound on all such pairwise 
conditional probabilities. This allows us to rank order significant patterns in terms 
the value of $e_0$ at which the pattern (which has a given number of repetitions) is 
no longer significant. The second interesting feature of the method follows from this 
structure of our composite null hypothesis. Since we now include many models of 
interdependent neurons (as long as all the relevant conditional probabilities are less 
than $e_0$), rejecting such a null hypothesis is intuitively more satisfying. 
When we declare a pattern such as $A\stackrel{T_1}{\rightarrow}B\stackrel{T_2}{\rightarrow}C$ as significant, we can 
conclude that a spike by $A$ has a ``strong'' influence in eliciting a spike from $B$ with delay $T_1$ and a spike from $C$ 
after a further delay of $T_2$. Here ``strong'' would denote that the relevant conditional probability is greater than $e_0$. Thus, our idea of casting the null hypothesis in terms of a bound on conditional probabilities allows for a richer 
level of analysis.

\begin{flushleft}{\bf Computational considerations}\end{flushleft}
We have given a simple test of statistical significance for deciding whether or not 
to reject the null (under a given confidence level) based on the counts calculated through 
simple multi-point correlations. As said earlier, the motivation is that such 
correlations are what are presently used for detecting such patterns. 

At this point one may wonder whether there is any need for the test of significance that 
we presented, given that we formulate our null hypothesis in terms of conditional 
probabilities. The correlations counts $f_{AB}$ as defined here would directly lead 
to an estimate of the conditional probability, $e(B|A,T)$. Hence, one can estimate the conditional probability and check whether it is less than $e_0$. While it is true that we can directly get an estimate of the conditional probability, 
to get the required confidence intervals on the estimate, we once again need to use 
similar kind of assumptions as here and hence, theoretically, the testing procedure 
is not irrelevant. But there are other reasons why this approach is better than 
estimating all conditional probabilities. 

First, our test will directly give the threshold needed for the count, given any 
pattern. Thus, we need not actually obtain the true correlation count which is required 
if we want to estimate the conditional probability. We only need to 
ascertain whether a pattern occurs more than some number of times. Many of the algorithms 
for detecting patterns use the correlations in this way and it leads to better 
computational efficiency. 

There is a second and more interesting reason why our approach could be beneficial. In general, 
obtaining correlation counts or ascertaining whether a pattern occurs a given number 
of times is computationally intensive. If we want to look for long patterns, the number 
of candidate patterns increases exponentially and, furthermore, the multi-point correlation is difficult to compute. However, there may be other more appealing ways to count what 
may be called the frequency of a pattern. The correlation count we considered here counts 
{\em all} occurrences of the relevant pattern. Suppose we want to count only those occurrences 
such that the time span of one occurrence does not overlap with that of any other 
occurrence. Let us call such occurrences {\em non-overlapped} occurrences of the pattern. 
(This means, e.g., if the spike sequence is $A A B B$, then we count only 
one occurrence rather than two). There are very efficient algorithms based on 
data mining techniques for obtaining all patterns whose counts in terms of maximum possible 
number of non-overlapped occurrences are above a 
threshold \cite{PSU2008}. These algorithms are also computationally efficient in 
discovering very long patterns involving more than ten neurons \cite{PSU2008}. 
It appears possible to extend this type of statistical significance analysis to 
such counts also. Also, even when finding correlations, it is possible to tackle the combinatorial explosion in candidates (when we are looking for long patterns) by using similar data mining methods if we can put a bound on the count and decide that we are interested in only those patterns above this count. We will be addressing these issues in our future work. It is in terms of such generality that the approach presented here is interesting. 

\begin{flushleft}{\bf Strengths and weaknesses of the method}\end{flushleft}
The strength of the approach is that we can accommodate dependence between neurons and 
conclude that some pattern is significant only if it represents strong influences 
among the set of neurons. Since this strength of influence is controlled by a 
parameter in the null hypothesis, we can rank order different significant patterns 
by varying this parameter. 
 
However, the weakness of the specific test proposed here is that we need to assume that the 
first neuron in the chain is Poisson. The assumption was needed to conclude that the pattern counts would be Poisson. In most of the currently available methods, one also assumes Poisson processes in the null hypothesis. 
In general, if the variations in the rate of firing of a neuron 
are small (which would be the case if all synapses into the neuron are very weak) then 
the Poisson assumption is likely to be a good approximation. Thus, the assumption is not restrictive 
if we know that some neuron is necessarily the first in a chain. However, it is not always possible to have such knowledge. In spite of the assumption of Poissonness of the first neuron, we feel that the approach presented here is interesting and useful.

\begin{flushleft}{\bf Summary}\end{flushleft}
In this paper we suggested an analytical method to assess the statistical 
significance of sequential firing patterns with constant delays between successive 
neurons. The main methods of detecting such patterns depend on multi-point 
correlations. Our method can be used to find thresholds on such correlation counts 
for deciding on the significance of the patterns. Our main motivation is to have 
a method that can rank order significant patterns in terms of the strength of influence 
among the neurons constituting the pattern. For this we suggested that the influence of 
$A$ on $B$ can be denoted in terms of the conditional probability of $B$ firing after a prescribed delay given that $A$ has fired. Our compound null hypothesis is then stated in terms of an upper bound on all such pairwise conditional probabilities. This upper bound is a parameter of the null hypothesis and by varying it we can compare 
different significant patterns in terms of the strength they represent. This feature 
is very novel in relation to the current methods of significance analysis. Another 
important consequence of our approach is that the null hypothesis now admits many 
models of interdependent neurons also in addition to the usual case of independence. 
Hence our approach to significance analysis is more general. 

Through extensive simulation experiments we demonstrated the effectiveness of the 
method. The method is seen to work well and is seen to be able to rank order different 
patterns in terms of their strengths even when our assumption in the theoretical 
analysis, namely that the first neuron in the chain is Poisson, is not valid. 

The method presented can assess significance of sequential firing patterns only when 
the underlying influences are excitatory. This is because the significance test is 
stated in terms of a lower bound on the correlation count. Using a similar null 
hypothesis where we assume that the conditional probability is much smaller than 
the case under independence, it may be possible to find how low the correlation count 
should be for us to conclude that there are significant inhibitory influences. This 
needs further investigation.  

As we have pointed out there are some weaknesses in the approach. One is the assumption 
that the first neuron in the sequence is Poisson. The other is the computational problems 
involved in finding correlation counts when one wants to detect interactions among a large group of neurons. There are some efficient algorithms based on data mining techniques which find somewhat different counts but are computationally very efficient for discovering patterns involving large numbers of neurons \cite{PSU2008}. We will be addressing the issue of extending the analysis presented here to such counts in our future work. 

\clearpage

\section{Appendix A}
\label{sec:appA}
\begin{table}[htbp]
\centering
\begin{tabular}{|c|c|}
\hline
Symbol & Meaning \\ \hline \hline
\parbox{0.75 in}{$A-Y$} & \parbox{5 in}{neurons} \\ \hline
\parbox{0.75  in}{$T (T_1,T_2,\ldots)$} & \parbox{5 in}{delay between neurons} \\ \hline
\parbox{0.75  in}{$\Delta T$} & \parbox{5 in}{time resolution of measurement / update rate of simulator} \\ \hline
\parbox{0.75  in}{$f_{AB}$} & \parbox{5 in}{correlation integral (\# of spikes of $A$ followed by $B$ with a delay of $T$)} \\ \hline
\parbox{0.75  in}{$I_x(t)$} & \parbox{5 in}{indicator variable for a spike in a time interval of $\Delta T$ centered around $t$} \\ \hline
\parbox{0.75  in}{$e(B|A,T)$} & \parbox{5 in} {conditional probability that $B$ will fire in a time interval $[T-{\Delta T}/2, \ T+{\Delta T}/2]$ given that $A$ has fired at time 0}\\ \hline
\parbox{0.75  in}{$e_0$} & \parbox{5 in}{user chosen bound on conditional probability for null hypothesis ($e(Y|X,T)\leq e_0$)} \\ \hline
\parbox{0.75  in}{$L$} & \parbox{5 in}{total time duration of data} \\ \hline
\parbox{0.75  in}{$N_A(L)$} & \parbox{5 in}{random variable for \# of spikes by $A$ in data} \\ \hline
\parbox{0.75  in}{$S_{AB}$} & \parbox{5 in}{random variable representing correlation integral $f_{AB}$} \\ \hline
\parbox{0.75  in}{$X_i$} & \parbox{5 in}{independent and identically distributed 0-1 random variables} \\ \hline
\parbox{0.75  in}{$p$} & \parbox{5 in}{probability that $X_i=1$} \\ \hline
\parbox{0.75  in}{$\lambda_A$} & \parbox{5 in}{rate of Poisson process for spiking of $A$} \\ \hline
\parbox{0.75  in}{$\lambda_S$} & \parbox{5 in}{expectation of random variable $S_{AB}$ $(pL\lambda_A)$} \\ \hline
\parbox{0.75  in}{$Z$} & \parbox{5 in}{Poisson random variable with mean $\lambda_Z$} \\ \hline
\parbox{0.75  in}{$\alpha$} & \parbox{5 in}{Type I error in null hypothesis} \\ \hline
\parbox{0.75  in}{$M$} & \parbox{5 in}{significance threshold (smallest number satisfying $P[Z>M]\leq \alpha$)} \\ \hline
\parbox{0.75  in}{$n$} & \parbox{5 in}{number of neurons in chain or pattern} \\ \hline
\end{tabular}
\label{tab:tab1}
\end{table} 

\section{Appendix B}
\label{sec:appB}
\begin{flushleft} {\bf Simulation model} \end{flushleft}
Here we describe the simulator used for generating spike data from a 
network of interconnected neurons. The spiking of each
neuron is an inhomogeneous Poisson process whose rate of firing is
updated at time intervals of $\Delta T$. The neurons are interconnected by synapses and each synapse is characterized by
a delay (which is in integral multiples of $\Delta T$) and a weight which is a real number. All neurons also have a refractory period. The rate of the Poisson process is varied with time as follows:
\begin{equation}
\lambda_j(k) = \frac{K_j}{1 + \exp{(-I_j(k) + d_j)}}
\label{eq:lambda-update}
\end{equation}
where $\lambda_j(k)$ is the firing rate of $j^{th}$ neuron at time $k \Delta T$, and $K_j, d_j$ are two parameters. $I_j(k)$ is the total input into $j^{th}$ neuron at time $k \Delta T$ and it is given by
\begin{equation}
I_j(k) = \sum_i O_i(k) w_{ij}
\label{eq:input}
\end{equation}
where $O_i(k)$ is the output of $i^{th}$ neuron (as seen by
the $j^{th}$ neuron) at time $k \Delta T$ and $w_{ij}$ is the weight of synapse from $i^{th}$ to $j^{th}$ neuron.
$O_i(k)$ is taken to be the number of spikes by the $i^{th}$ neuron in the time
interval $(\;(k-h_{ij}-1) \Delta T, \ (k-h_{ij}) \Delta T]$ where $h_{ij}$ represents the
synaptic delay (in units of $\Delta T$) for the synapse from $i$ to $j$.

We build the network in the following manner. The parameter $K_j$ is chosen based on the dynamic range of firing rates that we
need to span. The parameter $d_j$ is determined by specifying the background spiking rate. This is the firing rate of the neuron
under zero input. (We normally keep the same background firing rate for all neurons). Specifying this rate fixes $d_j$ by using (\ref{eq:lambda-update}). The network has many random interconnections with low weight values and a few strong interconnections with large weight values. For the random connections we connect each neuron to some percentage of all the other neurons randomly. The weight values for these random connections are uniformly distributed over a suitable range. We specify all weights in terms of the conditional probabilities they represent. Given a conditional probability, we first calculate the needed instantaneous firing rate so that probability of at least one spike in the $\Delta T$ interval is equal to the specified
conditional probability. Then using (\ref{eq:lambda-update}) and (\ref{eq:input}) we calculate the value of $w_{ij}$ needed so that the receiving neuron ($j$) reaches this instantaneous rate given that the sending neuron ($i$) spikes once
in the appropriate interval and assuming that input into the receiving neurons from
all other neurons is zero. In our simulations, we specify the range of random weight 
values as well as the values of strong weights in terms of the equivalent 
conditional probabilities. 

We then generate a spike train by simulating all
the inhomogeneous Poisson processes where rates are updated every $\Delta T$ time instants.
We also have a fixed refractory period for all neurons, so that once a neuron is fired we will not let it fire until the refractory period is over.

We note here that the background firing rate as well as the effective conditional
probabilities in our system would have some small random variations. As said above,
we fix $d_j$ so that on zero input the neuron would have the background firing rate.
However, all neurons would have synapses with randomly selected other neurons and
the weights of these synapses are also random. Hence, even in the absence of any
strong connections, the firing rates of different neurons keep fluctuating around the
background rate that is specified. Since we choose random weights from a zero mean
distribution, in an expected sense we can assume the input into a neuron to be
zero and hence the average rate of spiking would be the background rate specified.
We also note that the way we calculate the effective weight for a given conditional
probability is also approximate and we chose it for simplicity. If we specify
a conditional probability for the connection from $A$ to $B$, then, the method stated
earlier fixes the weight of connection so that the probability of
$B$ firing at least once in an appropriate interval given that $A$ has fired is equal
to this conditional probability {\em when all other input into $B$ is zero}. But since
$B$ would be getting small random input from other neurons also, the effective
conditional probability would also be fluctuating around the nominal value specified.
Further,  even if the random weights have zero mean, the fluctuations in the
conditional probability may not have zero mean due to the nonlinear sigmoidal
relationship in (\ref{eq:lambda-update}). The nominal conditional probability
value determines where we operate on this sigmoid curve and that determines
the bias in the excursions in conditional probability for equal fluctuations in either
directions in the random input into the neurons. We consider this as some more noise in the
system and have shown through simulation that our method of significance analysis is still effective. 

\section{Acknowledgements}
We thank Debprakash Patnaik for providing his Java implementation of the two-tape algorithm and Dr. Vijay Nair for helpful discussions.

\end{document}